\documentclass[onecolumn,  
showpacs,  
preprintnumbers,  
aps,  
prd,  
a4paper,  
nofootinbib,  
tightenlines,  
floats, floatfix  
]{revtex4}

\usepackage{amsmath}
\usepackage{amsfonts}
\usepackage{graphicx,color}

\newcommand{\bea}{\begin{eqnarray}}
\newcommand{\ena}{\end{eqnarray}}

\newcommand{\PP}{{\cal P}}

\begin{document}


\title{Primordial power spectrum versus extension parameters beyond the standard model}

\bigskip

\author{Zong-Kuan Guo}
\email{guozk@itp.ac.cn}
\affiliation{State Key Laboratory of Theoretical Physics, Institute of Theoretical Physics,
Chinese Academy of Sciences, P.O. Box 2735, Beijing 100190, China}

\author{Yuan-Zhong Zhang}
\email{zyz@itp.ac.cn}
\affiliation{State Key Laboratory of Theoretical Physics, Institute of Theoretical Physics,
Chinese Academy of Sciences, P.O. Box 2735, Beijing 100190, China}

\date{\today}

\begin{abstract}
We reconstruct the shape of the primordial power spectrum of
curvature perturbations in extended cosmological models,
including addition of massive neutrinos, extra relativistic
species or varying primordial helium abundance,
from the latest cosmic microwave background data from the
Wilkinson Microwave Anisotropy Probe, the Atacama Cosmology
Telescope and the South Pole Telescope.
We find that a scale-invariant primordial spectrum is disfavored by
the data at 95\% confidence level even in the presence of
massive neutrinos, however it can lie within the 95\% confidence region
if the effective number of relativistic species or the primordial
helium abundance is allowed to vary freely.
The constraints on the extension parameters from WMAP7+ACT+$H_0$+BAO,
are the total mass of neutrinos $\sum m_\nu < 0.48$ eV (95\% CL),
the effective number of relativistic species $N_{\rm eff}=4.50\pm0.81$
and the primordial helium abundance $Y_p=0.303\pm0.075$.
The constraints from WMAP7+SPT+$H_0$+BAO,
are $\sum m_\nu < 0.45$ eV (95\% CL), $N_{\rm eff}=3.86\pm0.63$
and $Y_p=0.277\pm0.050$.
\end{abstract}


\maketitle

\section{Introduction}

Measurements of temperature anisotropy in the cosmic microwave
background (CMB) provide us with a wealth of cosmological information.
Since 2001 the Wilkinson Microwave Anisotropy Probe (WMAP)
satellite has measured temperature anisotropy of the CMB up
to $l\sim 1200$ over the full sky~\cite{kom08,kom10}.
The ground-based telescopes such as the Atacama Cosmology
Telescope (ACT)~\cite{dun10} and the South Pole Telescope
(SPT)~\cite{kei11} have extended temperature measurements to
smaller angular scales $l\sim 3000$.
Measurements of the small-scale anisotropy of the CMB, in
combination with WMAP's measurements of the degree-scale
anisotropy, provide a powerful probe of early-universe physics.
Increasing the angular dynamic range of CMB measurements
can improve the constraints on the shape of the primordial power
spectrum of curvature perturbations~\cite{hlo11,guo11a,guo11b}.
Moreover, hight resolution measurements of the small-scale CMB
can limit the total mass of neutrinos, the effective number of
relativistic species and the primordial helium
abundance~\cite{dun10,kei11}.

Slow-roll inflationary models generically produce a nearly
scale-invariant primordial power spectrum of curvature perturbations.
Hence, measuring deviations from scale invariance of the
spectrum is a critical test of cosmological inflation.
For a power-law parameterization of the spectrum motivated by
inflationary models with featureless inflaton potentials,
the exact scale-invariant spectrum is excluded at more than 99\%
confidence level by the 7-year WMAP data~\cite{kom10}.
Other various parameterizations of the spectrum
have been considered: for example, a broken spectrum~\cite{bla03}
caused by an interruption of the inflaton potential~\cite{bar01},
a cutoff at large scales~\cite{efs03,bri06}
motivated by suppression of the lower multipoles in the CMB
anisotropies~\cite{nol08,cop10},
and more complicated shapes of the spectrum caused by
features in the inflaton potential~\cite{sta92}.
However, a strong theory prior on the form of the primordial
power spectrum could lead to misinterpretation and biases in
parameter determination.
Recently some model-independent approaches have been proposed to
reconstruct the shape of the primordial power spectrum, based on
linear interpolation~\cite{bri03},
cubic spline interpolation in log-log space~\cite{guo11a,guo11b},
a minimally-parametric reconstruction~\cite{sea05},
wavelet expansions~\cite{muk03}, principle component
analysis~\cite{hu03}, and a direct reconstruction via
deconvolution methods~\cite{kog03,sha03,toc04}.
In these works, the primordial power spectrum has been
constrained within the context of the minimal $\Lambda$CDM model.

On the other hand, recent CMB measurements by the ACT and
SPT experiments have revealed somewhat less fluctuation power
at small angular scales than expected in the standard
$\Lambda$CDM model.
Except for with the help of the primordial power spectrum with
a large negative running spectral index, such a small-scale
suppression in the CMB angular power spectrum can also be
explained by extra relativistic degree of freedom or additional
primordial helium abundance~\cite{dun10,kei11}.
Moreover, the addition of massive neutrinos can affect the CMB
anisotropies by changing the matter-to-radiation ratio around
the photon decoupling epoch~\cite{kom08}.

In this work, we reconstruct the shape of the primordial power
spectrum in the extended $\Lambda$CDM models, including
addition of massive neutrinos, extra relativistic species
or varying primordial helium abundance.
In order to reduce the degeneracy between the primordial power
spectrum and the extension parameters, we use the seven-year
WMAP data in combination with small-scale CMB data from the
ACT during its 2008 season and the SPT during 2008 and 2009,
and adopt two main astrophysical priors on the Hubble constant
($H_0$) measured from the
magnitude-redshift relation of low-$z$ Type Ia supernovae and
on the distance ratios of the comoving sound horizon to
the angular diameter distances from the Baryon Acoustic
Oscillation (BAO) in the distribution of galaxies.

This paper is organized as follows. In Section~\ref{sec2} we
describe the reconstruction method of the primordial power
spectrum and the extended models in which the power spectrum
is reconstructed.
In Section~\ref{sec3} we describe the data used in our analysis
and then present our results.
Section~\ref{sec4} is devoted to our conclusions.

\section{Beyond the standard model}
\label{sec2}

In standard $\Lambda$CDM model, the primordial power spectrum
is commonly parameterized as a power law,
there are three light neutrino species in the early Universe,
and big bang nucleosynthesis (BBN) took place with specific
predictions for primordial element abundances.
In this work, we consider simple extensions of the standard
$\Lambda$CDM model
including (i) general form of the primordial power spectrum,
(ii) addition of massive neutrinos, (iii) extra relativistic species,
(iv) varying primordial helium abundance.
Each type of model is in detail described below.

{\it Primordial power spectrum:}
The primordial power spectrum, $\PP(k)$, is related to the
angular power spectrum of the CMB by a radiative transfer
function $\Delta_l^X(k)$ via
\bea
C_l^{XY} = \int d\ln k \PP(k)\Delta_l^X(k)\Delta_l^Y(k),
\ena
where $X,Y$ denote the various temperature and polarization
modes. Given a specific ansatz for the primordial power spectrum
motivated by theoretical models, one can fit it to the data.
Alternatively, given our complete ignorance of the underlying
physics of the very early Universe, one can reconstruct
the shape of the primordial power spectrum from existing data.
In this paper, we use the method proposed in~\cite{guo11a}
to perform a reconstruction of the primordial
power spectrum in the extended $\Lambda$CDM models.
We divide the spectrum into three bins equally spaced
in logarithmic wavenumber between $0.0002$ Mpc$^{-1}$ and $0.2$ Mpc$^{-1}$.
We use a cubic spline interpolation to determine logarithmic
value of the spectrum between knots.
Outside of the wavenumber range we fix the slope of the spectrum
at the boundaries since the CMB data place only weak
constraints on them.

{\it Massive neutrinos:}
If the neutrinos are non-relativistic at the present epoch and
their mass eigenstates are degenerate, then the energy density
parameter, $\Omega_\nu$, can be written in terms of the total
mass of neutrinos as
\bea
\Omega_\nu h^2 \simeq \frac{\sum m_\nu}{94\,\rm{eV}},
\ena
which as a dark matter component in the Universe cannot be
negligible even for a neutrino mass as small as $m_\nu=0.07$ eV,
as established by the solar and atmospheric neutrino
oscillations experiments~\cite{gon10}.
Here, $h$ is the dimensionless Hubble parameter such that
$H_0=100h$ kms$^{-1}$Mpc$^{-1}$.
However, at the photon decoupling epoch neutrinos were still
relativistic as long as $\sum m_\nu < 1.8$ eV, and thus the
matter density at decoupling was actually smaller than a
naive extrapolation from the present value.
This leads to accelerating decay of gravitational potential
around the decoupling epoch.
Therefore, by changing the matter-to-radiation ratio the
massive neutrinos modify the CMB power spectrum~\cite{kom08}.
Detecting this effect on the CMB anisotropies allows us
to establish the absolute neutrino mass scale~\cite{ma95}.

{\it Extra relativistic species:}
In the standard model of particle physics there are three light
neutrino species. The total energy density of relativistic neutrinos
is related to the photon energy density, $\rho_\gamma$, via
\bea
\rho_\nu = \frac78 \left(\frac{4}{11}\right)^{4/3} N_{\rm{eff}}\,\rho_\gamma,
\ena
where $N_{\rm eff}$ is the effective number of neutrino species
that is used to parameterize the relativistic degree of freedom.
Three light neutrino species correspond to $N_{\rm eff}=3.046$,
which slightly exceeds 3 because electron-positron annihilation
at neutrino decoupling provides residual neutrino heating~\cite{man05}.

The addition of extra relativistic species increases the
expansion rate prior to and during the photon decoupling
epoch and thereby increases the angular scales of photon
diffusion, as explained in~\cite{hou11}.
Thus, increasing $N_{\rm eff}$ reduces fluctuation power
at small angular scales.
The effect from extra relativistic species not only produces a
nearly constant decrease in the angular power spectrum at
$l>200$ but also alters its shape at $l>200$~\cite{hou11}.

{\it Primordial helium abundance:}
In the standard BBN model, the predicted value of the primordial
helium abundance, $Y_p$, is generally related to the baryon
density and the effective number of relativistic species.
For the $\Lambda$CDM model, the uncertainty in the baryon density
leads to the $\sim0.02\%$ uncertainty in the helium abundance,
as $N_{\rm eff}$ is fixed to its standard value of 3.046.
Therefore, for CMB analysis the primordial helium abundance
is usually assumed to $Y_p=0.24$~\cite{kom08,kom10,dun10}.

Since the helium recombination happens much earlier than
the hydrogen recombination, the number density of free
electrons at the hydrogen recombination is given by
$n_e = n_b(1-Y_p)$ where $n_b$ is the baryon number density~\cite{hu95}.
A large $Y_p$ decreases the value of $n_e$ and thereby
increases the mean free path of photon.
This leads to lower fluctuation power in the damping tail of the CMB.
We can extend the BBN prior by promoting $Y_p$ to a free
parameter and then use the CMB data to probe the helium abundance.

To summarize, we consider the extended $\Lambda$CDM models
described by the parameters
\bea
\{\Omega_b h^2,\Omega_c h^2, \Theta_s,\tau,A_1,A_2,A_3,X\},
\ena
where $\Omega_b h^2$ and $\Omega_c h^2$ are the physical baryon and
cold dark matter densities relative to the critical density,
$\Theta_s$ is the ratio of the sound horizon to the angular diameter
distance at decoupling, $\tau$ is the reionization optical depth,
$A_i \equiv \ln [10^{10} \PP(k_i)]\,(i=1,2,3)$ are the logarithmic
values of the primordial power spectrum at knots $k_i$,
and $X=\sum m_\nu$, $N_{\rm eff}$ or $Y_p$ is the extension parameter.

\section{Cosmological constraints}
\label{sec3}

In this section, we first describe the CMB data from the WMAP
satellite, the ACT and SPT experiments, and astrophysical priors
used in our analysis.
We then combine these data sets to constrain the extended
$\Lambda$CDM models and present our results.

{\it Data:} We use the 7-year WMAP (WMAP7) data including the high-$l$
TT power spectrum in $l \leq 1200$ and TE power spectrum
in $l \leq 800$, and the low-$l$ temperature ($2 \leq l \leq 32$)
and polarization ($2\leq l \leq 23$) data.
We consider the Sunyaev-Zel'dovich (SZ) effect, in which CMB
photons scatter off hot electrons in clusters of galaxies.
Given a SZ template the effect is described by a SZ template amplitude
$A_{\rm SZ}$ as in the WMAP papers~\cite{kom08,kom10}.

We use the 148 GHz ACT data during its 2008 season.
The focus is on using the band powers in the multiple range
$1000 < l < 3000$ to improve constraints on primary cosmological
parameters.
Following~\cite{dun10} for computational efficiency
the CMB is set to zero above $l=4000$ where the contribution is
subdominant, less than 5\% of the total power.
To use the ACT likelihood described in~\cite{dun10}, aside from
$A_{\rm SZ}$ there are two more secondary parameters, $A_p$ and $A_c$.
The former is the total Poisson power from radio and
infrared point sources. The latter is the template amplitude of
the clustered power from infrared point sources.
We impose positivity priors on the three secondary parameters,
use the SZ template and the clustered source template provided by
the ACT likelihood package, and marginalize over these secondary
parameters to account for SZ and point source contamination.

We also use the 150 GHz SPT data during 2008 and 2009.
We focus on the temperature power spectrum binned in 40 band powers
in the multiple range $1000 < l < 3000$.
Following~\cite{kei11} Gaussian priors on the three
secondary parameters are adopted in our analysis.
The priors on the Poisson power, cluster power and SZ power
are $A_p=19.3 \pm 3.5\; \mu {\rm K}^2$,
$A_c=5.0 \pm 2.5\; \mu {\rm K}^2$ and
$A_{\rm SZ}=5.5 \pm 3.0\; \mu {\rm K}^2$, respectively.
Moreover, we require these secondary parameters to be positive.
As discussed in~\cite{kei11}, constraints on cosmological
parameters do not depend strongly on these priors.
The foreground terms are used only when calculating the SPT
likelihood; they are not used when calculating the WMAP likelihood.

To improve constraints on cosmological parameters we follow
the methodology described in~\cite{kom08,kom10} to consider the
addition of distance measurements from astrophysical observations,
on the angular diameter distances measured from BAO
in the distribution of galaxies,
and on the Hubble constant measured from the magnitude-redshift
relation of low-$z$ Type Ia supernovae.
The Gaussian priors on the distance ratios are
$r_s/D_V(z=0.2)=0.1905 \pm 0.0061$ and $r_s/D_V(z=0.35)=0.1097 \pm 0.0036$,
derived from the two-degree field galaxy redshift survey and
the sloan digital sky survey data~\cite{per09}.
Here $r_s$ is the comoving sound horizon size at the baryon
drag epoch and $D_V$ is the effective distance measure for
angular diameter distance.
The Gaussian prior on the Hubble constant is
$H_0=74.2\pm 3.6$ km s$^{-1}$ Mpc$^{-1}$, derived from
the magnitude-redshift relation of 240 low-$z$ Type Ia supernovae
at $z<0.1$~\cite{rie09}. The error includes both statistical
and systematical errors.

\begin{figure}[!hbt]
\begin{center}
\includegraphics[width=75mm]{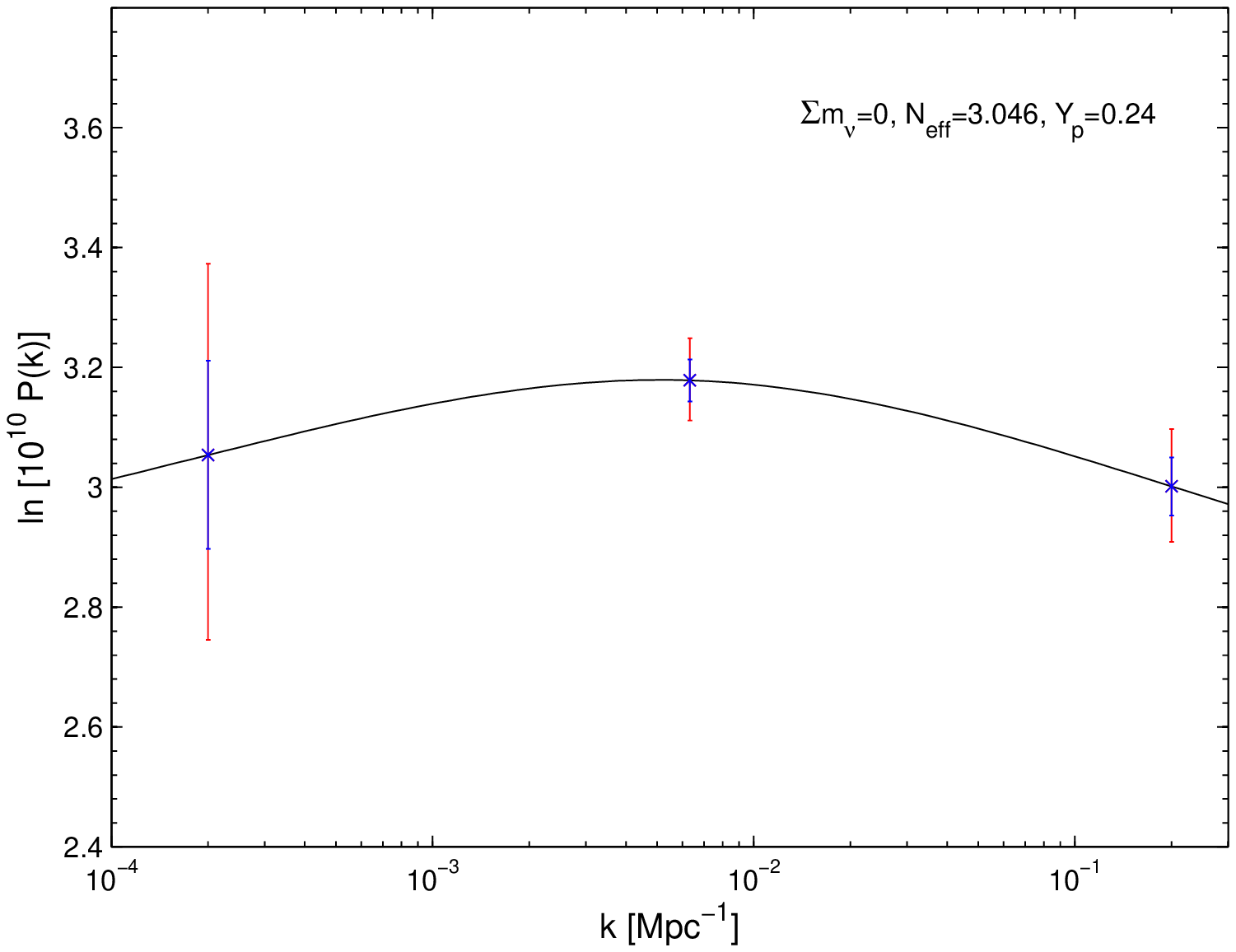}
\includegraphics[width=75mm]{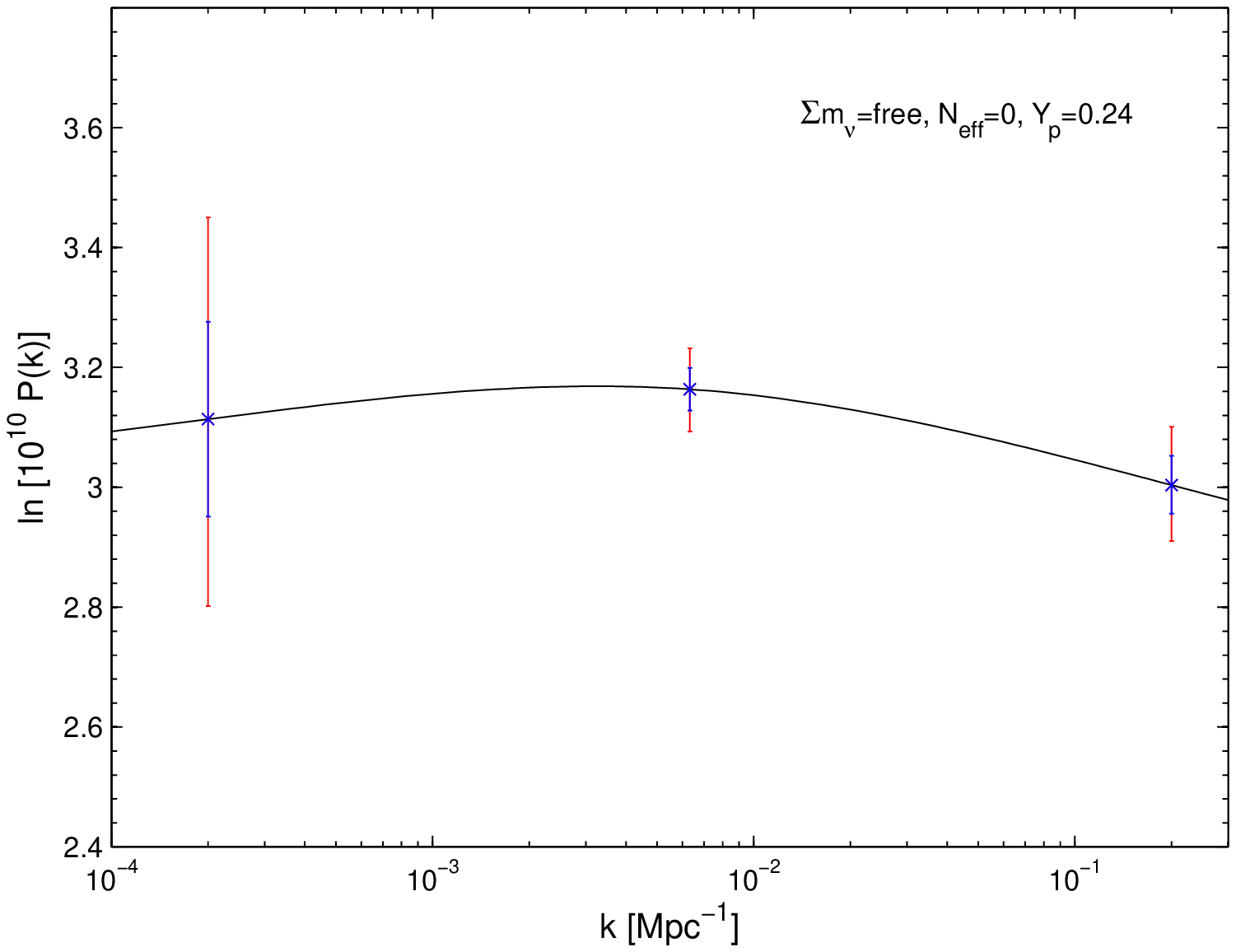}
\includegraphics[width=75mm]{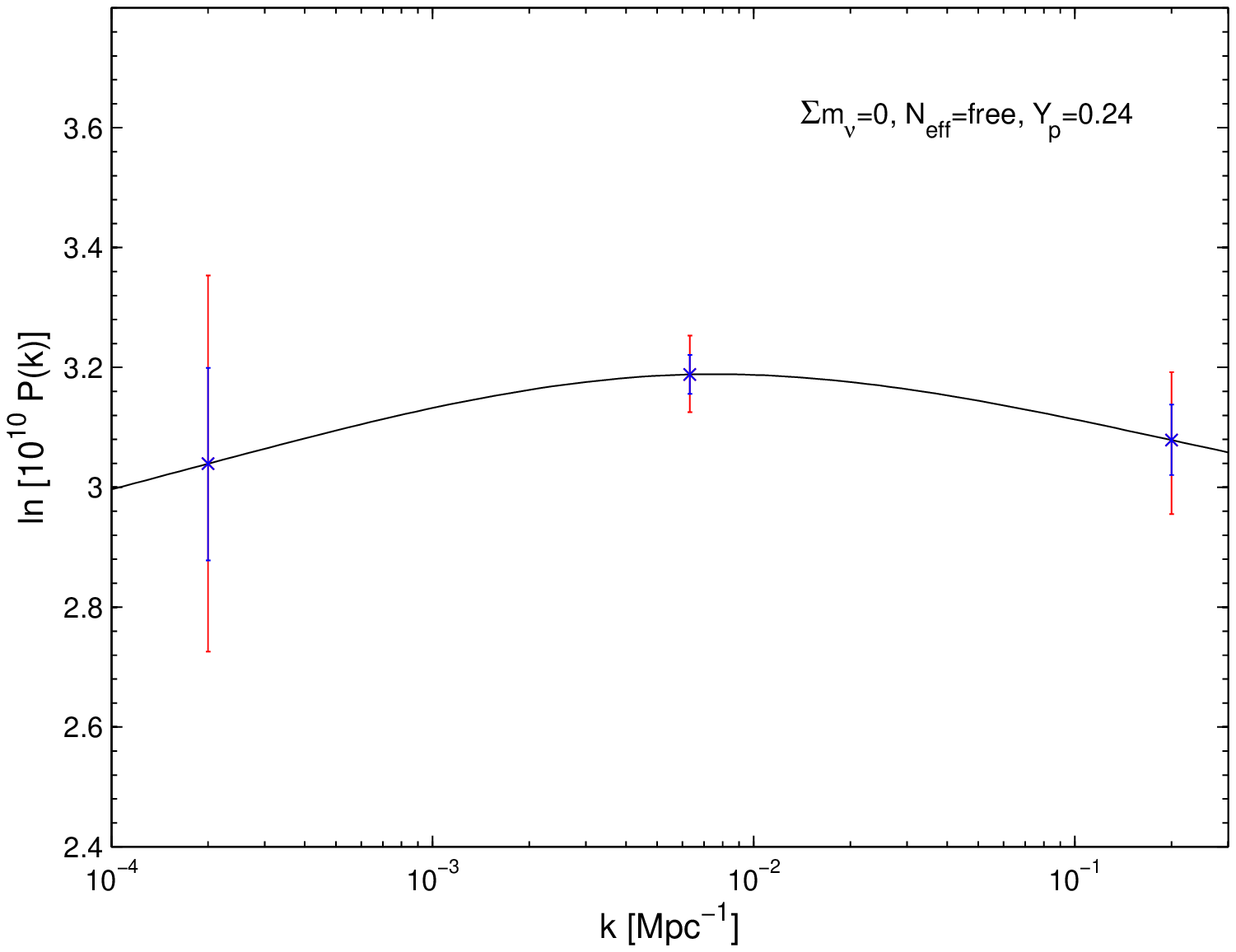}
\includegraphics[width=75mm]{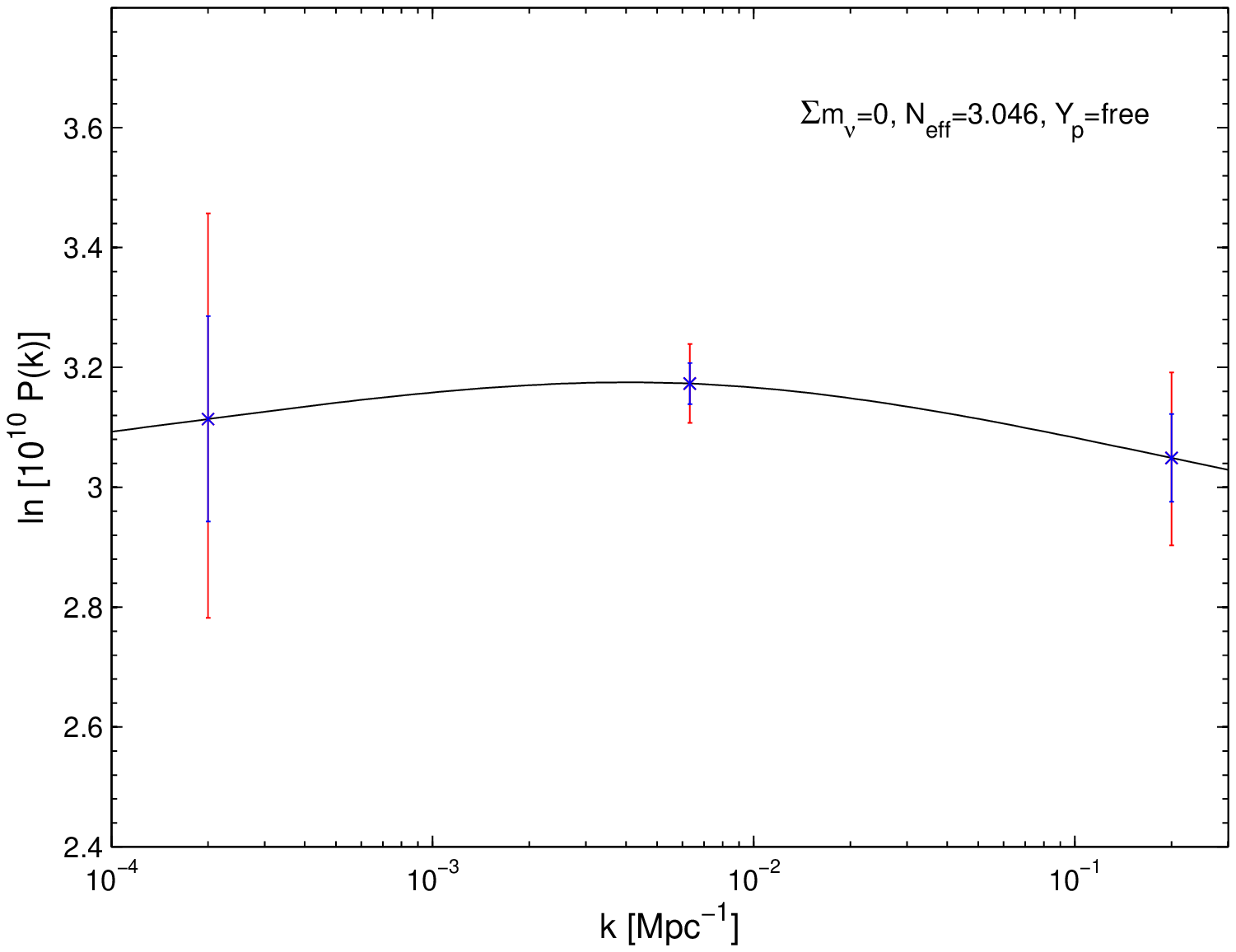}
\caption{Reconstruction of the primordial power spectrum of
curvature perturbations, $\PP(k)$, from WMAP7+ACT+$H_0$+BAO,
assuming $\sum m_\nu = 0$, $N_{\rm eff}=3.046$ and $Y_p=0.24$
(top-left panel);
$\sum m_\nu = {\rm free}$, $N_{\rm eff}=0$ and $Y_p=0.24$
(top-right panel);
$\sum m_\nu = 0$, $N_{\rm eff}={\rm free}$ and $Y_p=0.24$
(bottom-left panel);
$\sum m_\nu = 0$, $N_{\rm eff}=3.046$ and $Y_p={\rm free}$
(bottom-right panel).}
\label{fig-act}
\end{center}
\end{figure}

\begin{figure}[!hbt]
\begin{center}
\includegraphics[width=120mm]{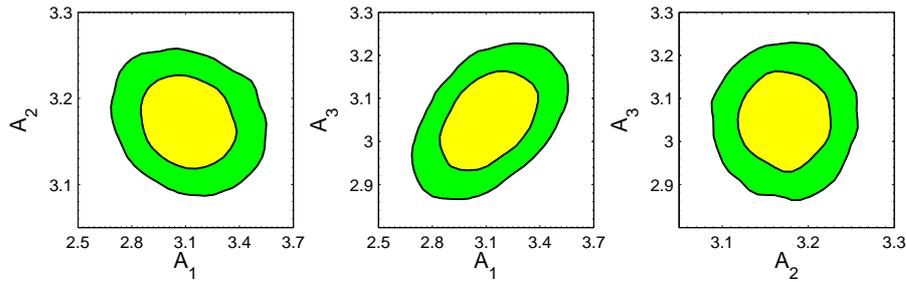}
\caption{Two-dimensional joint marginalized constraints
(68\% and 95\% C.L.) on the primordial power spectrum parameters,
$A_1$, $A_2$ and $A_3$, derived from WMAP7+ACT+$H_0$+BAO.
}
\label{fig-corr}
\end{center}
\end{figure}

{\it Results:} We use a modified version of the publicly
available CosmoMC package to sample the parameter space
by means of Monte Carlo Markov Chains technique~\cite{lew02}.
We also use the WMAP likelihood, the ACT likelihood and
the SPT likelihood code for parameter estimation.
Figure~\ref{fig-act} shows the reconstructed primordial power
spectrum of curvature perturbations in the standard
$\Lambda$CDM model (top-left panel) and non-standard models
with massive neutrinos (top-right panel), with varying
relativistic species (bottom-left panel) and a varying
primordial helium abundance (bottom-right panel),
respectively, derived from WMAP7+$H_0$+BAO combined with
the ACT small-scale CMB data.
We can see that in the standard model the scale-invariant
spectrum is disfavored by the data at 95\% confidence level.
Including massive neutrinos enhances the primordial spectrum
at low-$k$ but suppresses it at middle-$k$ due to an
enhancement in the so-called early integrated Sachs-Wolfe effect.
Hence, the spectrum becomes flat at large scales.
However, there is no significant degeneracy between massive
neutrinos and the primordial power spectrum at small scales.
Therefore, the scale-invariant spectrum is marginally disfavored
by the data in the presence of massive neutrinos.
In the extended model with free number of relativistic species
(or free abundance of primordial helium), the scale-invariant
spectrum can lie well within the 95\% confidence region
because the considerable degeneracy between relativistic
species (or helium
abundance) and the spectrum results in large errors and
an enhancement of the spectrum at high-$k$.
Compared to varying relativistic species, varying helium
abundance leads to larger uncertainties of the primordial
spectrum at both low-$k$ and high-$k$.
Varying helium abundance affects the primordial spectrum
at small scales by modifying the recombination process.
It indirectly affects the primordial spectrum at large scales
due to the correlation between $A_1$ and $A_3$
as shown in Figure~\ref{fig-corr}.

\begin{table}[!htb]
\begin{center}
\begin{tabular}{lcccccc}
\hline
 & $\Lambda$CDM+$\PP(k)$ & $\Lambda$CDM+$\PP(k)$ & $\Lambda$CDM+$\PP(k)$ & $\Lambda$CDM+$\PP(k)$ \\
 &              & + $\sum m_\nu$ & + $N_{\rm eff}$ & + $Y_p$ \\
\hline
$A_1$ & $3.054\pm0.159$ & $3.114\pm0.165$ & $3.039\pm0.162$ & $3.114\pm0.172$ \\
$A_2$ & $3.179\pm0.035$ & $3.164\pm0.036$ & $3.188\pm0.033$ & $3.173\pm0.034$ \\
$A_3$ & $3.002\pm0.049$ & $3.004\pm0.049$ & $3.079\pm0.060$ & $3.049\pm0.073$ \\
$\sum m_\nu$ & 0 & $<0.48\;{\rm eV}$ & 0 & 0 \\
$N_{\rm eff}$ & 3.046 & 0 & $ 4.50\pm0.81$ & 3.046 \\
$Y_p$ & 0.24 & 0.24 & 0.24 & $0.303\pm0.075$ \\
\hline
\end{tabular}
\end{center}
\caption{Mean values and marginalized 68\% confidence level
for the primordial power spectrum parameters and extension
parameters, derived from WMAP7+ACT+$H_0$+BAO.
For the total mass of neutrinos, the 95\% upper limit is given.}
\label{tab-act}
\end{table}

In Table~\ref{tab-act} we list the estimated values of
the primordial power spectrum as well as the extension
parameters beyond the standard model.
For a flat $\Lambda$CDM model with a power-law power
spectrum, the WMAP7+$H_0$+BAO limit
on the total mass of neutrinos is $\sum m_\nu < 0.58$
eV (95\% CL)~\cite{kom10}.
In our model we find that the WMAP7+ACT+$H_0$+BAO limit is
$\sum m_\nu < 0.48$ eV (95\% CL).
The upper limit on the total mass of neutrinos is improved
by adding the small-scale ACT data.
With the WMAP7+ACT+$H_0$+BAO data combination, the effective
number of neutrino species is estimated to be
$N_{\rm eff}=4.50\pm0.81$ (68\% CL), which indicates that
a Universe with no relativistic neutrinos is excluded at
5.6$\sigma$ and the standard model with $N_{\rm eff}=3.046$
is consistent at 1.8$\sigma$.
This estimated value is consistent with the effective
number of relativistic species estimated in the power-law
$\Lambda$CDM model from the WMAP7+$H_0$+BAO data,
$N_{\rm eff}=4.34\pm0.88$ (68\% CL)~\cite{kom10},
and from the WMAP7+ACT+$H_0$+BAO data,
$N_{\rm eff}=4.56\pm0.75$ (68\% CL)~\cite{dun10}.
Therefore, the constraints on the effective number of
relativistic species do not depend strongly on additional
degree of freedom of the power spectrum.
For CMB analysis the primordial helium abundance is
ususally assumed to be $Y_p=0.24$.
When it is allowed to vary, we find $Y_p=0.303\pm0.075$
(68\% CL) using the WMAP7+ACT+$H_0$+BAO data.
This mean value is lower than predicted in the power-law
$\Lambda$CDM from the WMAP7+ACT data, $Y_p=0.313\pm0.044$
(68\% CL), because suppressed power spectrum of primordial
curvature perturbations at small scales contributes to
enhanced Silk damping of the CMB angular power spectrum.
However, the degeneracy between the power spectrum and the
helium abundance leads to larger uncertainties in $Y_p$
compared to the power-law spectrum.

\begin{figure}[!htb]
\begin{center}
\includegraphics[width=75mm]{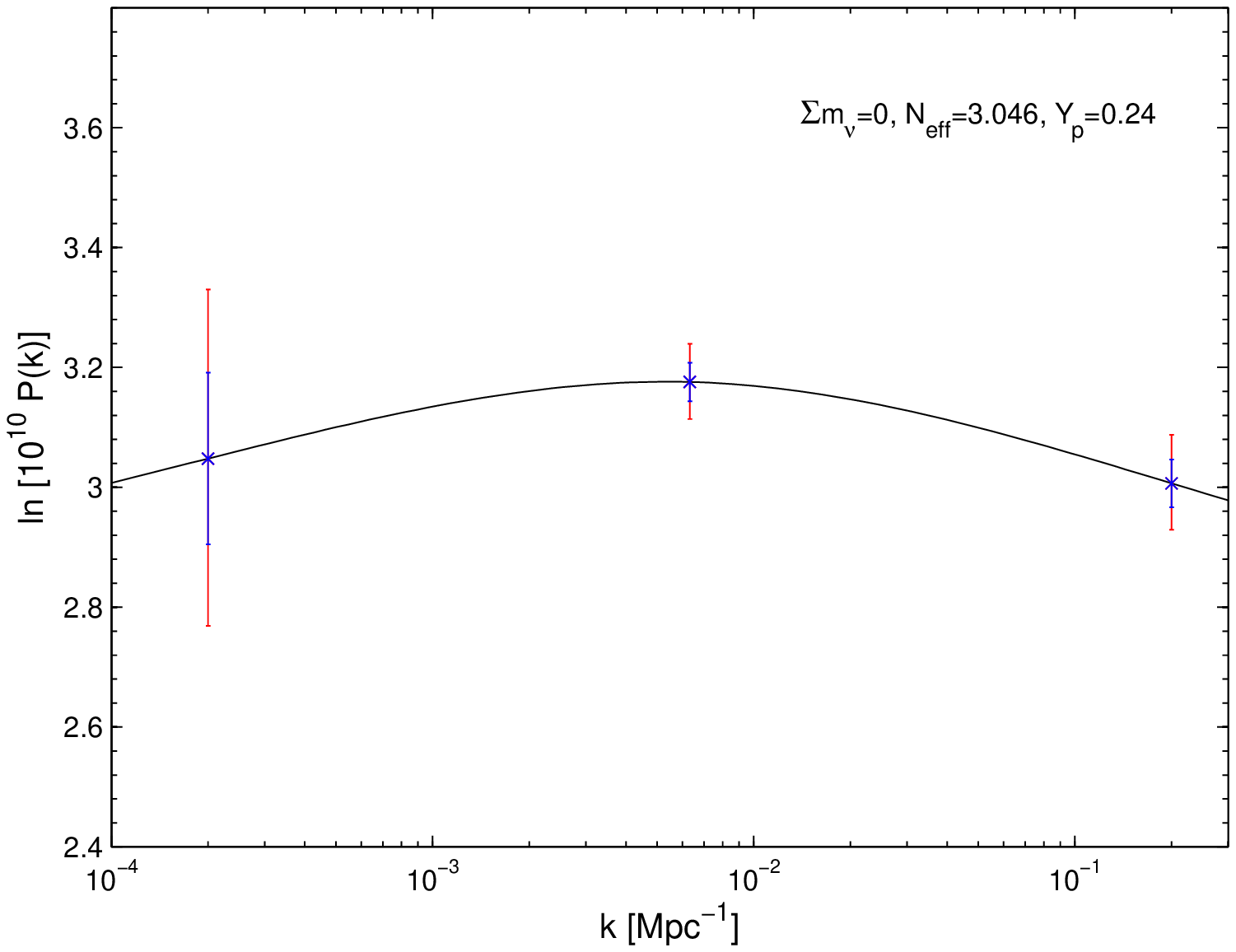}
\includegraphics[width=75mm]{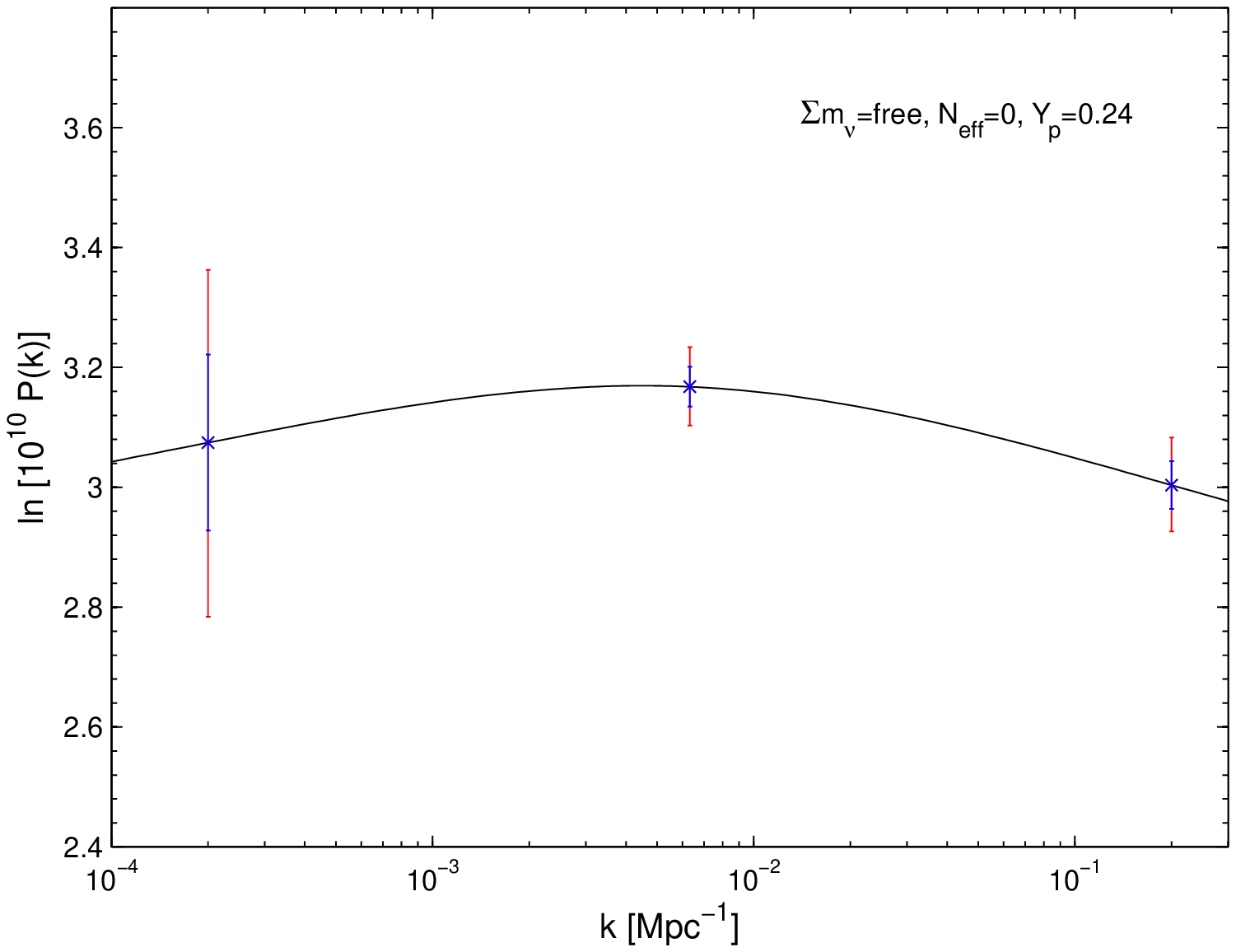}
\includegraphics[width=75mm]{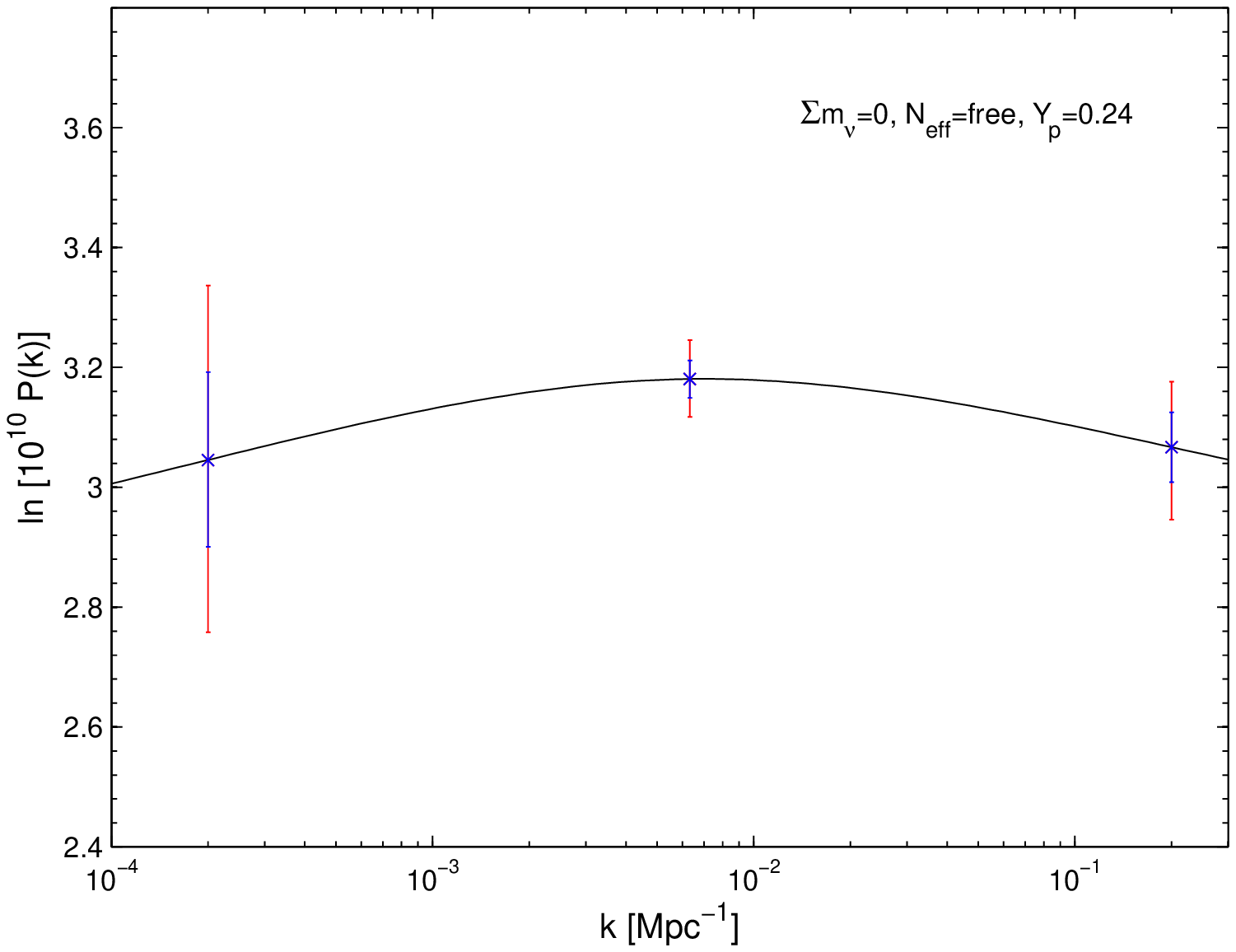}
\includegraphics[width=75mm]{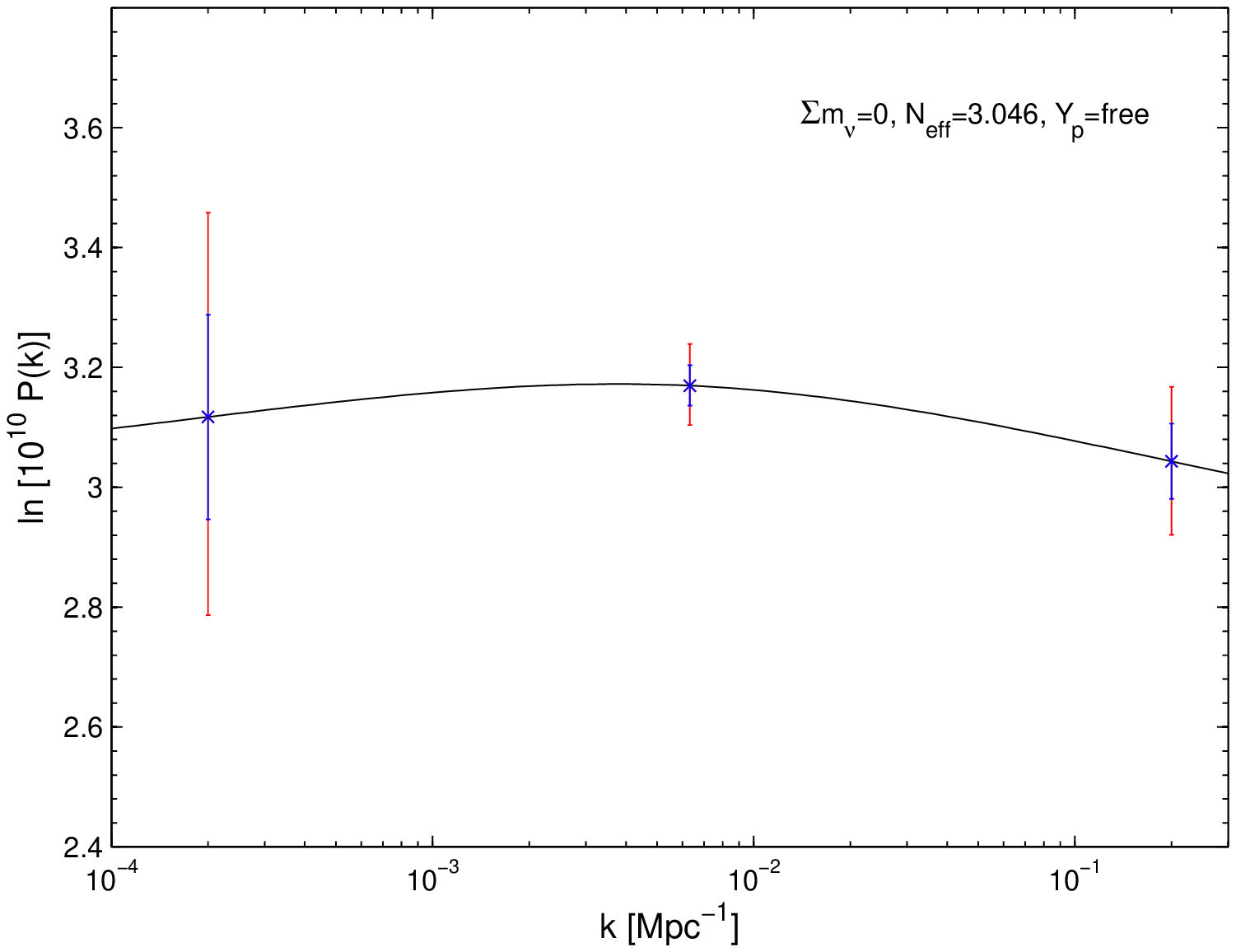}
\caption{Reconstruction of the primordial power spectrum of
curvature perturbations, $\PP(k)$, from WMAP7+SPT+$H_0$+BAO,
assuming $\sum m_\nu = 0$, $N_{\rm eff}=3.046$ and $Y_p=0.24$
(top-left panel);
$\sum m_\nu = {\rm free}$, $N_{\rm eff}=0$ and $Y_p=0.24$
(top-right panel);
$\sum m_\nu = 0$, $N_{\rm eff}={\rm free}$ and $Y_p=0.24$
(bottom-left panel);
$\sum m_\nu = 0$, $N_{\rm eff}=3.046$ and $Y_p={\rm free}$
(bottom-right panel).}
\label{fig-spt}
\end{center}
\end{figure}

\begin{table}[!htb]
\begin{center}
\begin{tabular}{lcccccc}
\hline
 & $\Lambda$CDM+$\PP(k)$ & $\Lambda$CDM+$\PP(k)$ & $\Lambda$CDM+$\PP(k)$ & $\Lambda$CDM+$\PP(k)$ \\
 &              & + $\sum m_\nu$ & + $N_{\rm eff}$ & + $Y_p$ \\
\hline
$A_1$ & $3.048\pm0.143$ & $3.074\pm0.147$ & $3.046\pm0.147$ & $3.117\pm0.171$ \\
$A_2$ & $3.176\pm0.032$ & $3.168\pm0.033$ & $3.181\pm0.032$ & $3.170\pm0.034$ \\
$A_3$ & $3.007\pm0.040$ & $3.004\pm0.040$ & $3.067\pm0.059$ & $3.043\pm0.063$ \\
$\sum m_\nu$ & 0 & $<0.45\;{\rm eV}$ & 0 & 0 \\
$N_{\rm eff}$ & 3.046 & 0 & $3.86\pm0.63$ & 3.046 \\
$Y_p$ & 0.24 & 0.24 & 0.24 & $0.277\pm0.050$ \\
\hline
\end{tabular}
\end{center}
\caption{Mean values and marginalized 68\% confidence level
for the primordial power spectrum parameters and extension
parameters, derived from WMAP7+SPT+$H_0$+BAO.
For the total mass of neutrinos, the 95\% upper limit is given.}
\label{tab-spt}
\end{table}

The constraints on the power spectrum and extension parameters
from the combination data of WMAP7+SPT+$H_0$+BAO are shown
in Figure~\ref{fig-spt} and listed in Table~\ref{tab-spt}.
Compared to the ACT data, the addition of the SPT data
reduces uncertainties in the power spectrum at small scales.
The scale-invariant spectrum is disfavored at 2$\sigma$ even
in the extended model with massive neutrinos.
The conclusion that the scale-invariant spectrum is not
excluded at 2$\sigma$ confidence level if either the effective
number of relativistic species or the helium abundance
is allowed to vary, is confirmed by using the WMAP7+SPT+$H_0$+BAO data.
The combination data constrain the total mass of neutrinos
to be $\sum m_\nu < 0.45$ eV (95\% CL), which is almost consistent
with the result listed in Table~\ref{tab-act}.
We find that the WMAP7+SPT+$H_0$+BAO limit on the effective number
of relativistic species is $N_{\rm eff}=3.86\pm0.63$ (68\% CL).
In the power-law model, the constraint improves to be
$N_{\rm eff}=3.86\pm0.42$ (68\% CL)~\cite{kei11}.
Compared to the ACT data, the SPT data prefers to a small
effective number of relativistic species.
If the helium abundance is free to vary, we find the
constraint is $Y_p=0.277\pm0.050$ (68\% CL) from WMAP7+SPT+$H_0$+BAO.
This mean value is lower than the result in the power-law model
presented in~\cite{kei11}, $Y_p=0.300\pm0.030$ (68\% CL).
Moreover, compared to the ACT data, the SPT data prefers to
a low helium abundance.

\section{Conclusions}
\label{sec4}

It is known that addition of massive neutrinos can influence
the CMB spectrum by changing the matter-to-radiation ratio around
the photon decoupling epoch, that extra relativistic species
can suppress the CMB damping tail by increasing the expansion
rate prior to and during the decoupling epoch, and that
excessive helium abundance can also suppress the CMB power at
small angular scales by decreasing the number density of free
electrons at the hydrogen recombination.
Therefore, the extension parameters can mimic the shape of the
primordial power spectrum and meanwhile the general form of the
primordial power spectrum can affect the constraints on the
extension parameters.

In such extended $\Lambda$CDM models, we have reconstructed
the shape of the primordial power spectrum of curvature
perturbations by using the cubic spline interpolation
method in log-log space.
To reduce the degeneracy between the primordial power spectrum
and the extension parameters, we use the CMB data from WMAP7+ACT
and WMAP7+SPT in combination with measurements of the Hubble
constant and the BAO feature.
Even if massive neutrinos are included in the early Universe,
the scale-invariant spectrum is disfavored by the combined
data at 95\% confidence level.
If the effective number of relativistic species (or the primordial
helium abundance) is allowed to vary, then the scale-invariant
spectrum can lie within the $2\sigma$ bound because of the considerable
degeneracy between $A_3$ and $N_{\rm eff}$ (or $Y_p$).
Moreover, we found that the two data sets yield nearly the same
limit on the total mass of neutrinos.
Compared to the SPT data, the ACT data mildly prefers to a large
effective number of relativistic species or a large primordial
helium abundance.

The contribution from tensor modes was considered when testing
for the shape of the primordial power spectrum in~\cite{guo11a}.
Under the assumption of a scale-invariant tensor spectrum,
including tensor modes can suppress the spectrum at large scales.
In this case the reconstructed spectrum deviates from
the scale-invariant one at 95\% confidence level.

\acknowledgments
We thank Y.Y.Y. Wong for useful discussions.
Our numerical analysis was performed on the Lenovo DeepComp 7000 supercomputer in SCCAS.
This work is partially supported by the project of Knowledge Innovation
Program of Chinese Academy of Science, NSFC under Grant No. 11175225,
and National Basic Research Program of China under Grant No:2010CB832805.
We used CosmoMC and CAMB.
We also acknowledge the use of WMAP data, ACT data and SPT data.

\end{document}